\documentclass[conference]{IEEEtran}
\IEEEoverridecommandlockouts

\usepackage{cite}
\usepackage{amsmath,amssymb,amsfonts}
\usepackage{algorithmic}
\usepackage{graphicx}
\usepackage{textcomp}
\usepackage{xcolor}
\usepackage{url}
\usepackage{booktabs}%
\usepackage[perpage]{footmisc}
\usepackage{amsmath}
\usepackage{rotating}
\usepackage{hyperref}
\usepackage{subcaption}
\usepackage{caption}
\usepackage{float} 
\usepackage{multirow}
\usepackage[T1]{fontenc}

\begin{document}

\title{MambaCPU: Enhanced Correlation Mining with State Space Models for CPU Performance Prediction\\
\thanks{Intel.}
}

\author{\IEEEauthorblockN{1\textsuperscript{st} Xiaoman Liu}
\IEEEauthorblockA{\textit{Data Center and Artificial Intelligence (DCAI)} \\
\textit{Intel Corporation}\\
Shanghai, China \\
xiaoman.liu@intel.com \\
\href{https://xiaoman-liu.com/}{Personal Homepage}}
}


\maketitle

\begin{abstract}
	Forecasting CPU performance, which involves estimating performance scores based on hardware characteristics during operation, is crucial for computational system design and resource management. This research field currently faces two primary challenges. First, the diversity of CPU products and the specialized nature of hardware characteristics make real-world data collection difficult. Second, existing approaches, whether reliant on hardware simulation models or machine learning, suffer from significant drawbacks, such as lengthy simulation cycles, low prediction accuracy, and neglect of characteristic correlations. To address these issues, we first gathered, preprocessed, and standardized historical data from the 4th Generation Intel Xeon Scalable Processors across various benchmark suites to create a new dataset named PerfCastDB. Subsequently, we developed a novel network, MambaCPU (MaC), as the baseline model for the PerfCastDB dataset. This model employs the mamba structure to explore global dependencies and correlations among multiple characteristics. The use of intra- and inter-group attention mechanisms further refines correlations within and between characteristic groups. These techniques enhance MaC's capability to analyze and mine complex multivariate correlations. Comparative experiments on the PerfCastDB dataset demonstrate that MaC surpasses existing methods, confirming its effectiveness. Additionally, we have open-sourced part of the dataset and the MaC code at \url{https://github.com/xiaoman-liu/MaC} to facilitate further research..
\end{abstract}


\section{Introduction}

CPU performance prediction based on hardware characteristics is crucial for optimizing computational system design and resource management. It plays a pivotal role in tasks ranging from hardware design to resource allocation, as it allows for forecasting a CPU performance under various operational conditions. This capability is indispensable not only for CPU manufacturers, who seek to streamline the design and prototyping process, but also for consumers and enterprises that need to select the most suitable hardware for their specific workloads. In summary, CPU performance prediction possesses substantial theoretical research value and practical application significance, representing one of the most valuable research areas.

Numerous studies have focused on understanding and predicting computer system performance. Early work \cite{ein1987attributes, saavedra1996analysis, nussbaum2001modeling, lee2007illustrative} employed statistical and sampling methods to analyze computer performance. Subsequently, research shifted towards using machine learning methods \cite{hamerly2006using, lin2013application, malakar2018benchmarking, mankodi2020evaluating, tousi2022comparative} for CPU performance prediction. The explosive growth of deep learning in text \cite{vaswani2017attention}, speech \cite{hinton2012deep}, and image recognition \cite{he2016deep} has introduced new avenues for performance prediction \cite{dean2018new}. Dibyendu et al. \cite{das2018specnet} developed the deep neural network SpecNet, achieving high prediction accuracy. Yu Wang et al. \cite{wang2019predicting} demonstrated that deep neural network models significantly outperform traditional linear models in benchmark performance prediction. Michael et al. \cite{siek2023benchmarking} utilized a long short-term memory (LSTM) model for CPU and GPU performance prediction. 

CPU performance prediction remains a challenging field due to several key limitations. First, the collection of real-world data is hindered by the sheer variety of available CPU products, each with highly specialized and diverse hardware characteristics. As a result, the field lacks a unified dataset that can comprehensively cover different hardware configurations and benchmarks. Without such a standardized dataset, it becomes difficult to compare and evaluate the effectiveness of different prediction models across varying CPU architectures. Second, the previous approaches, ranging from hardware simulation models to machine learning-based methods, suffer from notable drawbacks. Hardware simulation models are often computationally expensive, requiring lengthy test cycles that limit their practicality for real-time applications. The machine learning methods offer faster predictions but often struggle with low accuracy and an inability to fully capture the complex correlations between different hardware characteristics. These methods typically treat the features as independent or fail to exploit the intricate dependencies between them, which are critical for accurate performance prediction.

According to the above analysis, we make contributions from two aspects to promote the development of the CPU performance prediction field. First, we organize a novel dataset for further research. Concretely, we collect the historical CPU benchmark data of the 4th Generation Intel\textsuperscript{\textregistered} Xeon\textsuperscript{\textregistered} Scalable Processors, including data samples with 83-dimensional the hardware characteristics and the 1-dimensional corresponding performance prediction scores under different benchmark suites. Following the data cleaning, data standardization and feature engineering processes, we can generate standard data instances in the dataset. As a result, we organize a novel CPU performance prediction dataset called PerfCastDB, each instance contains $35$ hardware characteristics, and $1$ ground truth prediction scores under $6$ testing suites. For better understanding of the organized data, we provide a sub-benchmark sample at the link \href{https://github.com/xiaoman-liu/NCPP/blob/main/data/raw/SPR/train_data.csv}{Intel Sapphire Rapids sample}. Second, we present MaC, a state-of-the-art model designed to uncover and leverage the complex dependencies between multiple hardware characteristics. The MaC model is built on the Mamba structure \cite{gu2023mamba,dao2024transformers,zhu2024vision,yu2024mambaout}, which is tailored to mine global dependencies across multivariate data. 

In general, the contributions of this article can be summarized as:
\begin{itemize}
	\item We organize a novel dataset called PerfCastDB, which suits for the CPU performance prediction task. This dataset offers comprehensive coverage of hardware characteristics and performance benchmarks, providing a solid foundation for future research in CPU performance prediction
	\item We propose a novel network MambaCPU(MaC) under the PerfCastDB dataset. MaC is built on the Mamba structure, which can leverage the complex dependencies and global correlations between multiple hardware characteristics. The intra- and inter-group attention mechanisms are introduced to refine the group-wise correlations. The comprehensively utilization of the character correlations ensures the prediction accuracy and interpretability of the MaC model.
	\item We compare the proposed MaC with several traditional approaches. The experimental results show that MaC significantly outperforms existing approaches, which evaluates the effectiveness of MaC.
	
\end{itemize}

\section{Dataset Organization}
Data was collected from Sapphire Rapids (SPR), the 4th Generation Intel\textsuperscript{\textregistered} Xeon\textsuperscript{\textregistered} Scalable Processors based on Intel 7 technology \cite{Products57}, from September 27, 2022, to October 27, 2023. This dataset encompasses various stock-keeping units within the SPR product line and employs multiple testing suites, including SPECrate2017, Memory Latency Checker (MLC), Stream, and High Performance Conjugate Gradients (HPCG). The SPECrate suite \cite{SPECCPU90} is divided into "SPECrate2017\_int\_base" and "SPECrate2017\_fp\_base." MLC \cite{IntelMe18} provides metrics on local and cross-socket latencies and bandwidth, offering 9 levels of latency and 9 types of bandwidth data. Stream \cite{STREAMBe96} evaluates memory efficiency under various scenarios, while HPCG \cite{dongarra2016new} assesses the performance and efficiency of high-performance computing systems.

We standardize the dataset through five key stages: outlier cleaning, multi-output conversion, feature trimming, feature expansion, and normalization/tokenization, resulting in a clean and high-quality dataset for model training. In outlier cleaning, we apply a z-score filtering method to remove data points with z-scores exceeding a threshold ($|z| > 3$), where $z = (x - \mu) / \sigma$. Multi-output conversion consolidates six types of benchmark performance data into a unified 11-dimensional vector, reducing redundancy and improving training efficiency. Feature trimming involves removing redundant and non-significant hardware characteristics to enhance generalizability. Feature expansion enriches memory-related features by extracting detailed specifications using identifiers like "DIMM.PartNo" from manufacturers such as Samsung \cite{ModuleDR99}, Hynix \cite{Technica69}, and Micron \cite{SearchRe73}. Finally, normalization and tokenization convert categorical features into numerical tokens and scale numerical features consistently, enhancing training efficiency and prediction accuracy. For detailed data formats, refer to our open-source code \cite{MACdata1}.

\section{Method}

\begin{figure*}[htbp]
	\centering
	\includegraphics[width=1\textwidth]{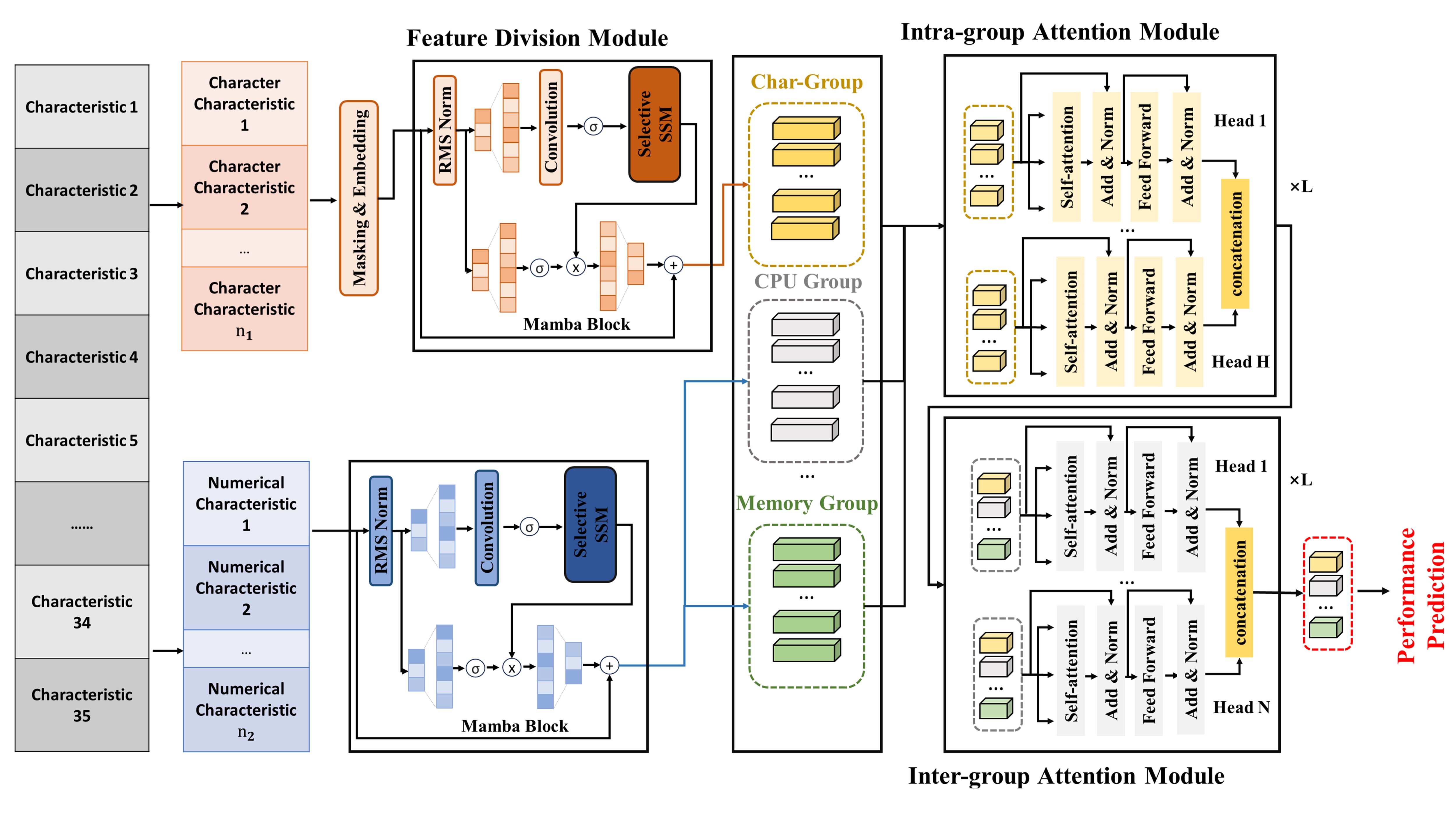}
	\caption{MaC Architecture: Feature Extraction from Character and Numerical Inputs Using a Selective state Space with Mamba Block and Multiple Attention Mechanisms}
	\captionsetup{font=small}
	\label{fig:model}
\end{figure*}

\subsection{State Space Model}
State space methods provide a mathematical framework for modeling. In a linear time-invariant system with input \( x(t) \), state \( h(t) \), and output \( y(t) \), the system equations are expressed as:

\begin{equation}
	\small
	\begin{aligned}
	h'(t) &= Ah(t) + Bx(t), \\
	y(t) &= Ch(t).
	\end{aligned}
	\end{equation}
where \( A \) is the state matrix influencing the state rate of change, \( B \) is the input matrix affecting the state, and \( C \) is the output matrix determining the output from the state.
Mamba has discreted \( A \) and \( B \) into a structured state space model by a timescale parameter \(\Delta\), the tranfornation by zero-order hold (ZOH) can be defined as follows:

\begin{equation}
	\small
    \begin{aligned}
    \overline{\mathbf{A}} &= \exp(\Delta \mathbf{A}), \\
    \overline{\mathbf{B}} &= (\Delta \mathbf{A})^{-1} (\exp(\Delta \mathbf{A}) - \mathbf{I}) \cdot \Delta \mathbf{B}.
    \end{aligned}
    \end{equation}
Let $\overline{\mathbf{A}} = e^{\mathbf{A} \Delta}, \, \overline{\mathbf{B}} = (\Delta \mathbf{A})^{-1} (e^{\mathbf{A} \Delta} - \mathbf{I}) \Delta \mathbf{B}$, then the previous equations can be transformed into the following form:
\begin{equation}
	\small
\begin{aligned}
    h(t_{k+1}) &= \overline{\mathbf{A}} h(t_k) + \overline{\mathbf{B}} x(t_k) \\
    y(t_{k+1}) &= Ch(t_{k+1}).
\end{aligned}
\end{equation}
Finally, through a global convolution operation, the output can be calculated as follows:
\begin{equation}
	\small
    \begin{aligned}
    \overline{\mathbf{K}} &= (\overline{\mathbf{C}}\overline{\mathbf{B}}, \overline{\mathbf{C}}\overline{\mathbf{A}}\overline{\mathbf{B}}, \ldots, \overline{\mathbf{C}}\overline{\mathbf{A}}^k\overline{\mathbf{B}}, \ldots)\\
     \quad y(k) &= x \ast \overline{\mathbf{K}}
\end{aligned}
    \end{equation}

\subsection{Feature Division Module}
In this module, input in Figure.\ref{fig:model} features are categorized into numerical and character features. Mamba blocks are then used for feature extraction from both categories, which is defined as follows:
\begin{equation}
	\small
\begin{aligned}
    \text{RMS Norm:} \quad x_1^{(n)} &= \text{Proj}(\text{RMSNorm}(x^{(n)})) \\
    \text{Convolution:} \quad x_2^{(n)} &= \text{Conv}(x_1^{(n)}) \\
    \text{Selective SSM:} \quad x_3^{(n)} &= \sigma(\text{SSM}(x_2^{(n)})) \\
    \text{Projection:} \quad x_4^{(n)} &= \text{Proj}(x_3^{(n)}) \\
    \text{Multiplication:} \quad x_5^{(n)} &= x_4^{(n)} \cdot \sigma(x_1^{(n)}) \\
    \text{Addition:} \quad y^{(n)} &= x_5^{(n)} + x^{(n)}
    \end{aligned}
\end{equation}
where  \( x^{(n)} \) represents the \( n \)-th input, \(\text{Proj}\) denotes linear projection, \(\text{Conv}\) is convolution, \(\sigma\) is the activation function, \(\text{SSM}\) stands for the state space model, and \( y^{(n)} \) is the \( n \)-th final output.

\subsection{Group Attention Module}

Features from the Feature Division Module are categorized into Char, CPU, Memory, and Other groups for specialized processing. X is the attention input, the intra group attention mechanism can be expressed by: 
\begin{equation}
	\small
	\begin{aligned}
	\text{Self-Attention} &= \text{softmax}\left(\frac{Q^{(l)}(K^{(l)})^T}{\sqrt{d_k}}\right)V^{(l)}  \\
	\text{head}_i^{(l)} &= \text{Self-Attention}(Q_i^{(l)}, K_i^{(l)}, V_i^{(l)})\\
	\text{MultiHead}^{(l)} &= \text{Concat}(\text{head}_1^{(l)}, \ldots, \text{head}_h^{(l)})W_O^{(l)}
\end{aligned}
	\end{equation}
$W_Q^{(l)}$, $W_K^{(l)}$, and $W_V^{(l)}$ are trainable weight matrices for layer $l$, with $d_k$ as the embedding dimension. The outputs of the self-attention heads are concatenated and linearly transformed. The inter-group attention mechanism mirrors the intra-group mechanism, using the output of the intra-group attention module. The final output is fed into a dense layer with units equal to the number of benchmarks.

\subsection{Loss Function}
In this study, we utilize the huber loss instead of Mean Absolute Error (MAE) or Mean Squared Error (MSE) due to its robustness to outliers. The formula of huber loss can be expressed by:

\begin{equation}
	\small
	\text{huber loss} = \left\{\begin{array}{ll}
		\frac{1}{2} \times\left(y-\hat{y}\right)^{2}             & \text { for }|y-\hat{y}| \leq \delta \\
		\delta \times\left(|y-\hat{y}|-\frac{1}{2} \delta\right) & \text { otherwise }
	\end{array}\right.
	\label{eq:huber}
\end{equation}
where $y$ is the true value, $\hat{y}$ is the predicted value, and $\delta$ is a hyperparameter that determines the threshold for the transition between the MAE and MSE loss functions.

\section{Experiments}

\subsection{Data Splitting and Cross-Validation}

We collect data across six suites, dividing it into 60\% for training, 20\% for validation, and 20\% for testing, as detailed in Table \ref{tab:dataset}. To improve prediction, we fine-tuned MaC model parameters for each suite. Using 5-fold cross-validation, we based the final evaluation on the average results, which helps prevent overfitting and optimizes hyperparameters for better generalization.

\begin{table}[!htbp]
	\centering
	\renewcommand{\arraystretch}{0.9} 
	\caption{the data distribution for the training, validation, and testing sets on the perfcastdb dataset under different suites.}
	\begin{tabular}{lccc}
		\toprule
		\textbf{Suite Name}              & \textbf{TrainSet} & \textbf{ValSet} & \textbf{TestSet} \\ \midrule
		SPECrate2017 Integer base        & 816                    & 204                         & 254                   \\
		SPECrate2017 FP base             & 756                    & 189                         & 236                   \\
		Memory Latency Checker Latency   & 140                    & 35                          & 42                    \\
		Memory Latency Checker Bandwidth & 922                    & 230                         & 294                   \\
		Stream                           & 2858                   & 714                         & 897                   \\
		HPCG                             & 2815                   & 704                         & 886                   \\\bottomrule
	\end{tabular}
	\label{tab:dataset}
\end{table}

\subsection{Evaluation Metrics}
We utilize the MAE, MSE, and mean absolute percentage error (MAPE) as the evaluation metrics. These metrics can be defined as the following equations:
\begin{equation}
	\small
	\begin{aligned}
		\text{MAE} & = \frac{1}{n} \sum_{i=0}^n \left| x_i - \hat{x}_i \right| \\
		\text{MSE} & = \frac{1}{n} \sum_{i=1}^{n} (x_i - \hat{x}_i)^2 \\
		\text{MAPE} & = \frac{1}{n} \sum_{i=1}^{n} \left| \frac{x_i - \hat{x}_i}{x_i} \right|
	\end{aligned}
	\label{combined}
\end{equation}
where $x_i$ is the ground truth, $\hat{x}_i$ is the predicted value, and $n$ is the sample size. MAE quantifies average error in output units, MSE is sensitive to outliers, and MAPE offers percentage-based error interpretation.

\subsection{Implementation Details}
We conducted our experiments on NVIDIA GeForce RTX 4090 GPU using the Adam optimizer with an exponential decay learning rate strategy. Our initial learning rate was set at 0.001, utilizing a batch size of 1, and we trained the model over 300 epochs.



\subsection{Ablation Study}
In this subsection, we conduct the ablation study to demonstrate the effectiveness of each part in MaC. Specifically, the main components in MaC are mamba blocks and attention blocks. The results in Table. \ref{tab:atten_ab} indicate that the mamba-based MaC method, which combines intra-group and inter-group attention, performs best across all metrics.

\begin{table}[!ht]
	\centering
	\renewcommand{\arraystretch}{0.9} 
	\caption{Evaluating the impact of mamba block and inter-group \& intra-group attention mechanisms on model accuracy}
	\resizebox{0.5\textwidth}{!}{
	\begin{tabular}{lcccc}
		\toprule
		\textbf{Models} & \textbf{MAE} & \textbf{MAPE} & \textbf{MSE} & \textbf{Median SE} \\ 
		\midrule
		Mamba                 & 7.23         & 1.51\%        & 226.81        & 15.42              \\
		Mamba + intra-group attention& 8.63         & 1.70\%        & 349.33       & 24.49              \\
		MaC                   & \textbf{5.90}        & \textbf{1.18}\%       & \textbf{174.60}      & \textbf{10.59}              \\
		MaC w/o char-group & 6.51         & 1.31\%        & 184.71      & 12.83              \\
		MaC w/o mem-group & 6.01         & 1.21\%        & 121.52       & 12.09              \\
		MaC w/o cpu-group & 6.22         & 1.23\%        & 181.86       & 11.47              \\
		MaC w/o other-group & 7.23         & 1.43\%        & 201.94       & 15.32              \\
		\bottomrule
	\end{tabular}}
	\label{tab:atten_ab}
	
\end{table}

\subsection{Hyperparameters Study} \label{hyperparam}
In this subsection, we conduct the comparison experiments on different settings of important hyperparameters, focusing on the size of \(\Delta\) projection $S$, the number of attention layers $L$, and the weight of huber loss $\delta$ on MAE, MSE, Median SE and MAPE. Table. \ref{tab:hyperparam} presents the corresponding results on "SPECrate2017\_FP\_base" suite. The results show that the combination of proj=2, state=8, layer=3, head=4 and $\delta$=10 performs best across almost all metrics.

When proj=4 and state=8, the MSE reached a group minimum of 102.54, yet the MAE, Median SE, and MAPE were not minimized. This suggests that low MSE alone does not guarantee superior predictive performance, necessitating the consideration of detailed SE metrics. As shown, the Median SE for this setup exceeds that of proj=2. Additionally, SEs at the 75\%, 90\%, and 95\% percentiles were higher than those for proj=4, though these are not presented due to space constraints. This is due to MSE's sensitivity to outliers present in the test set. Therefore, in contexts like CPU performance prediction, where outliers exist, careful metric selection is essential, as MSE alone is inadequate.

\begin{table}[htbp]
	\centering
	\renewcommand{\arraystretch}{0.9}  
	\caption{Optimization of specrate2017 fp hyperparameters on MAE, MSE, Median SE, and MAPE. The parameters proj, state, layer, and head correspond to the expansion factor, state vector dimension size, number of attention layers, and number of attention heads, respectively.}
	\begin{tabular}{cccccc}
		\toprule
		\textbf{Hyperparameter} & \textbf{Value} & \textbf{MAE}  & \textbf{MSE}   & \textbf{Median SE} & \textbf{MAPE}   \\ \midrule
		\multirow{5}{*}{proj=S, state=8} & 1              & 6.79          & 186.47         & 15.20          & 1.38\%          \\
		~                       & \textbf{2}     & \textbf{5.90} & 174.60 & \textbf{10.59} & \textbf{1.18\%} \\
		~                       & 4              & 6.01          & \textbf{102.54}          & 12.88          & 1.21\%          \\
		~                       & 8              & 6.48          & 144.43         & 15.31          & 1.33\%          \\
		~                       & 16              & 6.40          & 196.72          & 14.74          & 1.30\%          \\ \midrule
		\multirow{5}{*}{proj=2, state=N}& 1     & 7.03 & 164.32 & 17.27 & 1.50\% \\
		~                       & 2              & 6.47          & 175.96         & 13.83          & 1.32\%          \\
		~                       & 4              & 6.82          & 169.95         & 17.86         & 1.44\%          \\
		~                       & \textbf{8}              & \textbf{5.90}          & 174.60         & \textbf{10.59}          & \textbf{1.18\%}        \\
		~                       & 16              & 6.15          & \textbf{106.86}         & 14.81          & 1.27\%          \\ \midrule
		\multirow{3}{*}{layer=L, head=4} & 1 & 7.17 & 319.77 & 13.68 & 1.45\% \\
		~ & 2 & 6.12 & 136.69 & 12.52 & 1.25\% \\
		~ &  \textbf{3} & \textbf{5.28} & \textbf{65.02} & \textbf{12.24} & \textbf{1.08}\% \\ \midrule
		\multirow{5}{*}{layer=3, head=H}& 		 1 & 6.17 & 115.57 & 15.11 & 1.27\% \\
		~ & 2 & 6.22 & 223.80 & \textbf{10.92} & 1.25\% \\
		~ & 3 & 5.88 & 89.13 & 13.55 & 1.19\% \\
		~ & \textbf{4} & \textbf{5.28} & \textbf{65.02} & 12.24 & \textbf{1.08}\% \\
		~ & 5 & 5.87 & 78.60 & 14.62 & 1.20\% \\ \midrule
		\multirow{3}{*}{$\delta$}    & 0.1            & 7.22          & 194.87         & 17.42          & 1.50\%          \\
		~                       & 1     & 6.95 & 204.39 & 15.58 & 1.41\% \\
		~                       & \textbf{10}            & \textbf{5.28} & \textbf{65.02} & \textbf{12.24} & \textbf{1.08}\%     \\ \bottomrule
	\end{tabular}
	\label{tab:hyperparam}
\end{table}



\subsection{Comparison of Prediction Performance}
We compare our model with various methods, including machine learning and deep learning techniques. Lasso and Ridge regression address multicollinearity, with ElasticNet combining their strengths. SVM performs well in high-dimensional spaces, while XGB handles missing data and provides feature importance. In deep learning, LSTM and GRU excel at learning long-term dependencies. Next, we present experimental results comparing the performance of our MaC model with these baseline methods. The corresponding results are shown in Table.\ref{tab:model_performance}. As a result, MaC achieves the best evaluation results on most suites, which directly reflects the superiority of it.

\begin{table}[!htbp]
	\centering
	\caption{Comparison performance across six benchmarks, including SPECrate2017 Integer base (S-Int), SPECrate2017 FP base (S-FP), Stream, HPCG, Memory Latency Checker Latency (MLC-L), Memory Latency Checker Bandwidth (MLC-B).}
	\renewcommand{\arraystretch}{0.9} 
	\begin{tabular}{l@{\hskip 0.1in}c@{\hskip 0.1in}c@{\hskip 0.1in}c@{\hskip 0.1in}c@{\hskip 0.1in}c@{\hskip 0.1in}c}
		\toprule
		\textbf{Models} & \textbf{S-Int} & \textbf{S-FP} & \textbf{Stream} & \textbf{HPCG} & \textbf{MLC-L} & \textbf{MLC-B} \\
		\midrule
		\multicolumn{7}{c}{\textbf{MAE \textdownarrow}} \\
		\midrule
		Lasso & 9.49 & 11.45 & 12.92 & 4.01 & 3.68 & 3.56 \\
		Ridge & 11.61 & 16.47 & 11.91 & 3.32 & 4.05 & 4.92 \\
		EN & 23.32 & 22.24 & 28.52 & 5.21 & 5.61 & 6.94 \\
		SVM & 21.53 & 19.06 & 9.31 & 3.13 & 4.18 & 3.52 \\
		XGB & 7.81 & 6.29 & 4.48 & 3.31 & 3.66 & \textbf{1.24} \\
		LSTM & 31.68 & 30.21 & 31.10 & 2.84 & 7.92 & 8.88 \\
		GRU & 9.70 & 8.47 & 4.17 & 5.42 & 3.45 & 8.89 \\
		MaC & \textbf{7.29} & \textbf{5.28} & \textbf{1.99} & \textbf{2.44} & \textbf{2.67} & 1.36 \\
		\midrule
		\multicolumn{7}{c}{\textbf{MAPE \textdownarrow}} \\
		\midrule
		Lasso & 2.42 & 2.50 & 3.20 & 5.74 & 3.49 & 4.63 \\
		Ridge & 2.86 & 3.73 & 2.93 & 4.61 & 3.28 & 5.02 \\
		EN & 5.94 & 4.70 & 7.03 & 7.87 & 5.68 & 8.78 \\
		SVM & 5.44 & 3.68 & 2.44 & 4.38 & 3.81 & 3.01 \\
		XGB & 1.99 & 1.32 & 1.08 & 4.30 & 3.05 & \textbf{0.87} \\
		LSTM & 7.80 & 6.13 & 7.68 & 3.78 & 7.39 & 9.41 \\
		GRU & 2.59 & 1.77 & 1.05 & 8.22 & 3.38 & 9.41 \\
		MaC & \textbf{1.89} & \textbf{1.08} & \textbf{0.48} & \textbf{3.10} & \textbf{2.81} & 1.09 \\
		\midrule
		\multicolumn{7}{c}{\textbf{MSE \textdownarrow}} \\
		\midrule
		Lasso & 339.86 & 471.17 & 365.63 & 33.27 & 32.42 & 48.80 \\
		Ridge & 620.16 & 995.37 & 446.59 & 31.03 & 40.73 & 814.41 \\
		EN & 1160.41 & 868.87 & 1137.69 & 66.36 & 59.45 & 170.98 \\
		SVM & 1520.52 & 902.13 & 322.29 & 21.16 & 54.64 & 149.03 \\
		XGB & \textbf{233.81} & 97.11 & 98.48 & 22.61 & 43.75 & \textbf{12.09} \\
		LSTM & 2027.33 & 1577.07 & 1378.09 & 12.86 & 113.81 & 377.25 \\
		GRU & 1133.63 & 164.96 & 403.15 & 73.49 & 41.68 & 377.15 \\
		MaC & 240.55 & \textbf{65.02} & \textbf{48.45} & \textbf{11.52} & \textbf{26.30} & 16.75 \\
		\bottomrule
	\end{tabular}
	\label{tab:model_performance}
\end{table}

\begin{figure}[htbp]
	\begin{minipage}[b]{0.23\textwidth}
		\includegraphics[width=\textwidth]{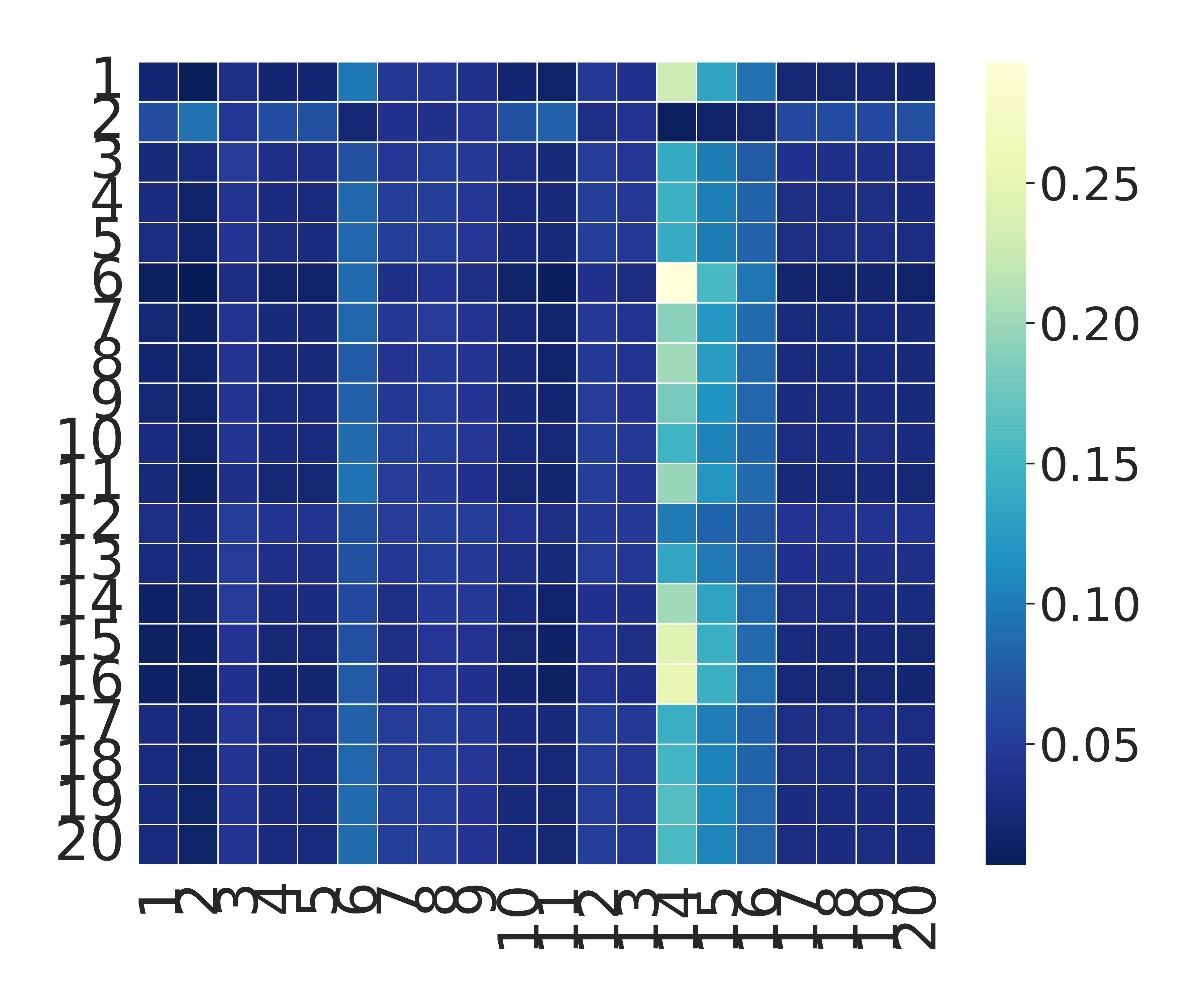} 
		\subcaption{CPU Group}
		\label{fig:1_cpu_matrix}
	\end{minipage}
	\begin{minipage}[b]{0.23\textwidth}
	\includegraphics[width=\textwidth]{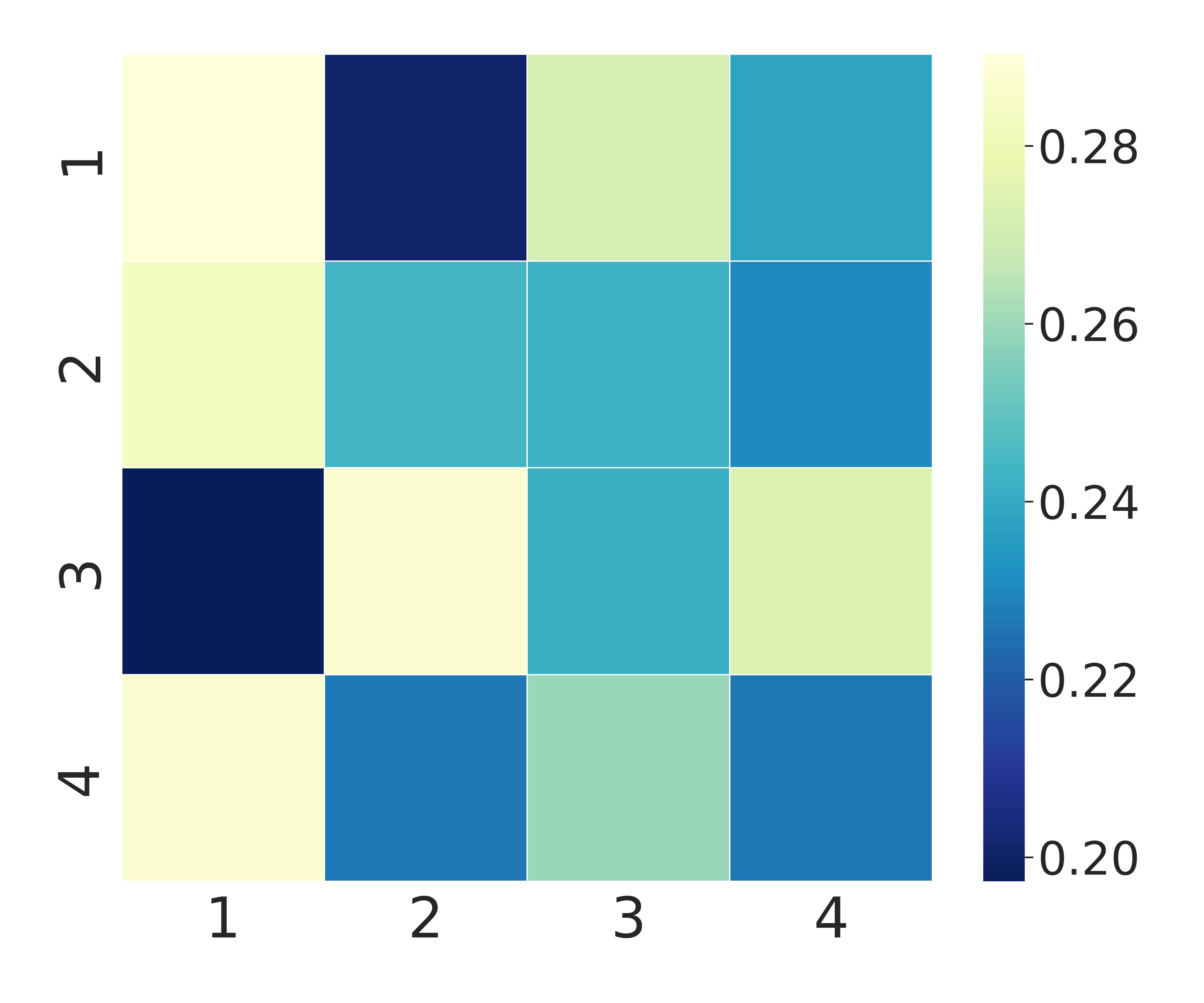}
	\subcaption{Inter Group}
	\label{fig:1_cross_matrix}
\end{minipage}
	\caption{Attention Matrix Visualization in CPU Intra-groups and Inter-group of the First Head for a SPEC CPU 2017 FP Data Sample}
	\label{fig:1_1_attention}
\end{figure}

In Figure.\ref{fig:1_1_attention}, we present the heatmaps for attention weights within and between feature groups. Figure.\ref{fig:1_cpu_matrix} illustrates the intra-group attention for the CPU group, comprising 20 feature sets, where lighter colors indicate stronger correlations. Figure.\ref{fig:1_cross_matrix} displays the inter-group attention across memory, CPU, char, and other groups. These patterns highlight the capability of the model to effectively capture and represent feature interactions, demonstrating the ability of attention mechanism to discern complex relationships.



\section{Conclusion}\label{sec13}
In this paper, we introduce PerfCastDB, a comprehensive benchmark dataset specifically designed for CPU performance prediction tasks. This dataset refines historical data to create a model-ready training set. We also present MaC as a baseline model, developed using mamba state-space equations combined with group attention mechanisms to enhance prediction accuracy. Extensive experiments demonstrate that MaC surpasses traditional methods, validating its effectiveness. Futhermore, employing diverse metrics is essential for accurately evaluating CPU performance prediction, particularly in the presence of outliers. We aim for our work to establish a robust foundation for future research in CPU performance prediction.


\bibliographystyle{IEEEtranS}
\bibliography{icassp}

\vspace{12pt}
\color{red}

\end{document}